# Development of a new UHV/XHV pressure standard (Cold Atom Vacuum Standard)

Julia Scherschligt[a], James A Fedchak*[,a], Daniel S Barker[b], Stephen Eckel[a], Nikolai Klimov[b], Constantinos Makrides[b], and Eite Tiesinga[a]

[a]National Institute of Standards and Technology, 100 Bureau Drive, Gaithersburg, MD 20899-8364, USA
[b]Joint Quantum Institute, University of Maryland, College Park, MD, 20742, USA

*corresponding author: james.fedchak@nist.gov, 1-301-975-8962

**Abstract**

The National Institute of Standards and Technology has recently begun a program to develop a primary pressure standard that is based on ultra-cold atoms, covering a pressure range of $1 \times 10^{-6}$ Pa to $1 \times 10^{-10}$ Pa and possibly lower. These pressures correspond to the entire ultra-high vacuum (UHV) range and extend into the extreme-high vacuum (XHV). This cold-atom vacuum standard (CAVS) is both a primary standard and absolute sensor of vacuum. The CAVS is based on the loss of cold, sensor atoms (such as the alkali-metal lithium) from a magnetic trap due to collisions with the background gas (primarily $H_2$) in the vacuum. The pressure is determined from a thermally-averaged collision cross section, which is a fundamental atomic property, and the measured loss rate. The CAVS is primary because it will use collision cross sections determined from *ab initio* calculations for the $Li + H_2$ system. Primary traceability is transferred to other systems of interest using sensitivity coefficients.

## 1. Introduction

The National Institute of Standards and Technology (NIST) has recently launched a program to develop a new type of standard for vacuum, which will employ laser-cooling and trapping of neutral atoms to make the Cold-Atom Vacuum Standard (CAVS). In the past several decades, the laser cooling and trapping of atoms has been a very active and exciting field, with numerous important discoveries. A variety of atom traps have been developed, including optical dipole traps and magnetic traps [1,2], which led to the realization of record low temperatures and Bose-Einstein condensates. Since the earliest days of neutral atom trapping it has been known that the background gas in the vacuum limits the lifetime of atoms in the trap. Indeed, this was pointed out in the first realization of a magnetic trap for neutral atoms [3]. Early rough estimates of trap lifetimes limited by background collisions give a lifetime of 100 s at a background pressure of $10^{-8}$ Pa [4] and this estimate is in reasonable agreement with observation. Several researchers have inverted the problem to use trap lifetime as a measurement of vacuum level [5-8]. However, a true absolute sensor of vacuum has not yet been realized. This is mostly due to the lack of knowledge of the collision cross section between the trapped ultra-cold atom and the ambient gas in the vacuum. Additionally, the systematic uncertainties associated with using an atom trap to determine vacuum level have not been thoroughly studied, particularly those associated with loss mechanisms (in a non-ideal trap) other than due to background collisions. Some studies have focused on the magneto-optical trap, where the collision physics in the non-conservative trap is complicated by the laser field [6]. The CAVS is based on a magnetic trap and an *ab initio* calculation of the loss-rate coefficient or collision cross section. A large part of our research effort will be to determine the loss-rate coefficients and relative sensitivity factors for various gases of interest.

Because the measured loss-rate of ultra-cold atoms from an ideal magnetic trap only depends on a fundamental atomic property (the collision cross section), such a device can be used as an absolute sensor and primary vacuum standard. It is our rigorous attention to theory—*ab initio* cross section calculations and accounting for non-ideal loss mechanisms—that differentiates CAVS from the earlier attempts at using trapped atoms to sense vacuum. Those did not rely on precise cross sections and less rigor was applied to accounting for non-ideal loss mechanisms. The CAVS is based on the following premise: ultra-cold (< 1 mK) "sensor" atoms held in a shallow magnetic trap ($E/k_B$ < 1 mK, where $k_B$ is the Boltzmann constant) will remain trapped until knocked out of the trap by collisions with ambient gas in the vacuum. The measured loss rate of sensor atoms from the magnetic trap determines the particle-number density of gas in the vacuum, $\rho_N$, which is related to the vacuum pressure by the ideal gas law, $p = \rho_N k_B T$ where $T$ is the temperature of the molecules in the vacuum. The CAVS is therefore both a primary standard, because it is based on a measurement of temperature, time, and fundamental atomic properties, and absolute sensor of vacuum. The CAVS will cover the pressure range of $10^{-6}$ Pa to $10^{-10}$ Pa and possibly lower. It bears emphasizing that this pressure range, from ultra-high vacuum (UHV) to extreme-high vacuum (XHV), is one in which no other existing primary pressure standard operates. Moreover, a goal of the nascent NIST program is to create a miniature portable version of the CAVS, thus creating a deployable primary vacuum standard based on fundamental quantum physics. The potential implications of such a device, which will belong to a class of quantum-based metrological tools seeking to recast the *System Internationale* (SI) units, are profound.

A new vision of metrology has evolved at NIST: the Quantum SI. Key elements to this vision correlate with the redefinition of the SI slated for 2018. Under the redefined SI, the base units will be tied to fundamental physical constants rather than an artifact or specific method of realization. Therefore, primary standards may be created using a variety of techniques. National Metrology Institutes (NMIs) or other calibration labs may create primary standards suited to their financial and uncertainty budget requirements. Within the Quantum SI paradigm, a standard must obey three "Laws." It must:

1) Be based upon unchanging, universal, fundamental constants or macroscopic quantum phenomena;

2) Provide a true value of the realization or measurement or no value at all; and

3) Have a known uncertainty that is characterized quantitatively and is fit for purpose [9].

The CAVS was conceived with the Quantum SI in mind: it is based on a fundamental atomic property and directly senses the gas in the vacuum. A prototype CAVS is being created with the lowest uncertainty possible for permanent installation and use at NIST for the calibration of transfer standards.

However, the advantages of the Quantum SI are not confined to the NMI. Portable devices, if based on the principles above, have clear advantages over traditional transfer standards. First, such devices never need to be returned to a NMI for recalibration (1st and 2nd law). Second, though the physics that underpins a Quantum SI device is absolute, its interrogation will still have some associated uncertainty, but these uncertainties will be well defined and quantitatively understood (3rd law). To realize these clear advantages for vacuum measurement, a portable version of the CAVS is also being created, allowing users to own and operate a fundamental vacuum standard for critical absolute measurements or calibrations.

NIST has a long history of laser cooling and trapping of neutral atoms, largely motivated by building better time standards or clocks. Indeed, four Nobel Prizes have been awarded to NIST researchers for developments in the cooling of neutral atoms and the construction of atomic clocks: Phillips in 1997, Cornell in 2001, Hall in 2005, and Wineland in 2012. The first magnetic trap from neutral atoms was realized at NIST in 1985 [3]. Over the decades, the field of ultra-cold atom physics has grown and applications of cold-atom science include atomic clocks, inertial sensing, gravity gradiometers, and quantum information and simulation. To bring these applications of ultra-cold atom technology from the laboratory to a more widespread user market will require a number of technical developments. Ultra-cold atom devices must operate in the UHV or lower, and the UHV environment must be maintained over the lifetime of the device. A vacuum of $10^{-6}$ Pa over a 1000 day period has been identified as one possible benchmark for these devices [10]. Portable ultra-cold atom devices for the Quantum SI must also be made small and robust enough to operate in harsh environments, such as a factory floor or in spaceflight instrumentation. A great deal of progress has been made in miniaturizing magneto-optical traps (MOT) and its variations for trapping ultra-cold atoms [10]. Most of the work has concentrated on trapping Rb atoms, and the miniaturization has focused on the MOT but not the supporting optics and photonics. The ultimate goal of this project is to create a robust portable CAVS as a first example of deployable cold-atom sensors, supporting the cold-atom science programs at NIST and other institutions.

This paper is structured as follows, we give an introduction to some of the fundamentals of atom trapping and laser cooling in section 2. The physics of the CAVS and its laboratory-scale prototype apparatus is described in section 3. Initial design of the CAVS began in earnest in 2016 and the first prototype is now under construction. Some of the details and design consideration are likely to change in future versions. Traceability of the NIST CAVS and the deployable CAVS will be discussed in section 3.

## 2. Fundamentals of Laser Cooling and Trapping of Neutral Atoms

The techniques for laser cooling and trapping of neutral atoms were developed in the 1980s. Here we provide a brief introduction; the interested reader is encouraged to find more detail in Ref. [11]. As stated above, the CAVS is based on trapping a cloud of cold atoms in a shallow, conservative trap (i.e., a trap where energy is conserved like an optical dipole trap or a magnetic trap) and measuring the rate at which background molecules eject cold atoms. The generation of a cold atomic cloud requires: (1) producing a vapor of atoms well above room temperature, (2) laser cooling those atoms and (3) placing those cold atoms into the conservative trap.

Not all atoms are easily laser cooled. The most commonly used atoms in laser cooling are alkali-metal atoms, which exist as a solid metal at room temperature. To vaporize atoms into the vacuum chamber for laser cooling generally requires heating this metal to create an appreciable partial vapor pressure (typically $10^{-6}$ Pa). For Li, this vapor pressure is only achieved at temperatures around 400 °C. For Cs, this vapor pressure is achieved naturally at around room temperature.

Once an appreciable vapor of hot alkali-metal atoms is made, laser cooling techniques can be applied to cool and trap those atoms. Three concepts are needed to understand laser cooling: (1) atomic energy levels are quantized, (2) light (or equivalently, a photon) carries momentum, and (3) the Doppler shift. Figure 1 depicts an idealized, stationary, two-state atom with energies $E_0$ and $E_1$. The resonant wavelength, $\lambda_0$, is related to $E = E_1 - E_0$ through Planck's constant, $h$, and the speed of light in vacuum, $c$, by $E = hc/\lambda_0$. If an atom at rest is exposed to this resonant light, it will absorb a photon (causing it to be excited to energy level $E_1$). After excitation, it will emit a photon in a *random* direction and decay back to the ground state. Therefore, after one such cycle, the atom receives a net momentum kick in the direction of propagation of the light. The lifetime of the excited state $\tau$ defines how often a momentum kick can be delivered to an atom, which is around 20 ns for alkali-metal atoms. This exchange of momentum between a photon and an atom at rest necessarily adds momentum to the atom. It is possible to employ the same physics to remove momentum from an atom in motion, thereby cooling it. To do this, one can exploit the Doppler effect to selectively change the velocity of atoms. In particular, if the atom is with velocity $v$, the photons in the laser beam will be Doppler shifted with wavelength equal to $\lambda' = \lambda \frac{c}{c-v}$ in the reference frame of the atom. By red-detuning the laser, one can ensure that only atoms with velocity toward the laser receive a momentum kick, slowing those atoms down. Atoms at rest and those moving in the same direction as the photons are unperturbed. This is the fundamental basis of Doppler-cooling, which can be used in three dimensions with multiple laser beams creating "optical molasses" to cool a cloud of atoms. The lowest temperature that can be reached in an optical molasses using this Doppler-cooling process is the Doppler temperature, $T_D = \frac{\hbar}{2\,k_B\tau}$. This temperature is of the order of 0.1 mK for alkali-metal atoms [11].

Laser cooling works quite well to reduce the temperature of atoms, but does nothing to confine them in space. To add spatial confinement, one exploits the Zeeman effect. For simplicity assume that the atom has a ground state with quantized angular momentum $J = 0$ and an excited state with $J = 1$. If a magnetic field of strength $B$ is applied, the atom's energy levels split according to $E = m_J \mu B$, where $\mu$ is the magnetic moment of the atom and $m_J$ is the projection of $J$ along $B$. If a magnetic field gradient is applied, the atom's energy levels depend linearly on its position. Figure 2 shows our model atom from Figure 1, but with a magnetic field $B$ that splits the $E_2$ level into three levels. The figure also shows two lasers, one with right-circular polarized light (driving transitions to the $m_J = +1$ state), and the other with left-circular polarized light (driving transitions to the $m_J = -1$ state). If a slow atom moves into the region near $z'$, it will become resonant with the rightward going beam and be pushed back toward the center of the trap. In this way, the atom can be trapped spatially and cooled using the Doppler cooling technique described above. This hybrid trap is known as a magneto-optical trap, and typically uses magnetic gradients of the order of $2 \times 10^{-3}$ T/cm.

Once the atoms are cooled and trapped in a 3D MOT, they can be transferred to a conservative trap that is purely magnetic, by switching off the lasers and adjusting the magnetic field such that it has gradients at least an order of magnitude larger. Additional techniques (for example, Sisyphus cooling [11]) will be applied to reduce the temperature of the atom cloud to well below 1 µK. In the ideal case, such cold atoms are only lost through collisions with background atoms and molecules. The trap is then said to have a lifetime $\tau_0$. A discussion of non-ideal conditions that either lead to additional atom loss or prevent atom loss is postponed until Section 4.

## 3. The CAVS apparatus

We will use the techniques from atomic physics described above to create the CAVS. Figure 3 shows a cutaway view of the prototype CAVS apparatus. There are two distinct regions in the CAVS: a source stage and a sensing stage, which are separated from each other by a constriction, which functions as a differential pumping tube. Because of the differential pumping tube's low conductance, there can be a significantly higher pressure in the source stage than than in the sensing stage.

In the source, alkali metal atoms (Rb or Li as described in detail later) are vaporized using *in-vacuo* resistively heated alkali sources. These effusion sources produce a stream of atoms that travel through a cold shroud (explained below). Using a MOT formed along two spatial dimensions (known as a 2D MOT), the atoms from the effusive source are redirected into a collimated beam. The beam is aligned with the differential pumping tube such that they pass through the differential pumping tube into the sensing chamber. Two precautions are taken to ensure that only atoms redirected by the 2D MOT enter the sensing stage. First, alkali-metal atoms emerging from the source are not directly aligned with the differential pumping tube. Second, a cooling shroud, located in the vacuum, surrounds the 2D MOT and is maintained between -30 °C and -50 °C. Most atoms not cooled by the 2D MOT must strike the cold shroud at least once in order to scatter into the differential pumping tube. Because the sticking coefficient for alkali-metal atoms is near unity on metal surfaces, the probability of rescattering into the sensing region is small. However, if a monolayer of Rb or Li accumulates on a surface, that coating will begin to emit atoms as an effusive source based on the saturated vapor pressure. To prevent reemission of atoms from the shroud, we maintain the shroud at -30 °C , where Rb has a saturation vapor pressure on the order of $10^{-9}$ Pa at 23 °C (The cooling shroud is not entirely necessary for Li because its vapor pressure is < $10^{-16}$ Pa at 23 °C) [12]. The flux of cold atoms emitted from the 2D MOT is expected to be on the order of $10^9$ atoms/s, and can be turned on and off simply by turning on and off the lasers of the 2D MOT.

The atoms from the 2D MOT then pass through the differential pumping tube into the sensing stage where they are captured into a 3D MOT. Here they are cooled to sub-millikelvin temperatures. The sensing stage consists of a pyrex glass cell surrounded by magnetic field coils. Six laser beams enter the cell along three primary axes. Once cooled, the laser beams are extinguished and the magnetic field configuration is changed from a quadrupole-type field necessary for the MOT [3] to a Ioffe-Pritchard type magnetic trap. The quadrupole field is produced by a subset of the magnetic coils used to produce the Ioffe-Prichard Trap [13,14]. The magnetic field arrangement of the Ioffe-Pritchard trap prevents the magnetic field from equaling zero within the trap which would otherwise lead to atom loss through so-called Majorana spin-flips [15]. The number of sensor atoms in the initially loaded magnetic trap is expected to be of the order of $10^6$, with a density of $10^{10}$ atoms/cm$^3$. Laser cooling is performed using circularly-polarized laser beams. For the 2D MOT, four beams pass through optical-quality viewports each attached to a vacuum nipple; for the 3D MOT, six beams pass through the glass cell. The laser beams are locked to the $D_2$ ($^2S_{1/2} \rightarrow {}^2P_{3/2}$) transition of the alkali-metal atoms, which is 780 nm for Rb and 671 nm for Li. The optical power in each beam is of the order of 10 mW. Not shown in Figure 3 are the magnetic-field coils that generate the fields required for the 2D MOT, 3D MOT, and magnetic trap.

Once the ultra-cold atoms are in the magnetic trap they are sensing the background gas in the vacuum. The CAVS stage is connected to the vacuum system of interest via a vacuum port on the CAVS cell, shown on the right in Figure 3. The density of sensor atoms in the magnetic trap is determined by absorption imaging of the ultra-cold atomic cloud in the trap. Collisions between the background gas molecules and the trapped sensor atoms will eject cold atoms out of the trap, necessarily reducing the density $\rho_S$ in the trap according to $\rho_S(t) = \rho_S(t=0)e^{-\Gamma t}$. The loss rate $\Gamma$ is extracted from the exponentially decaying density and is related to the background gas number density $\rho_N$ by

$$\rho_N = \Gamma / k_{loss} \qquad (1)$$

where $k_{loss} = <v\sigma>$ is the thermally averaged loss-rate coefficient, $v$ is the relative velocity, and $\sigma$ is the elastic collision cross section between the sensor atoms and the background molecules. The brackets denote a thermal average. Only the relative change in number density of the cold sensor atoms is necessary to determine $\Gamma$; knowledge of the absolute atom number is not.

Molecular hydrogen, $H_2$, is the most common residual gas in the UHV/XHV, and the CAVS will ultimately rely on precise knowledge of the Li + $H_2$ cross section. A precise value of the loss-rate coefficient for Li + $H_2$ will be determined from *ab-initio* quantum chemistry and scattering calculations. Thus the CAVS is a primary, Quantum SI-based, standard because it depends on a fundamental atomic property, $k_{loss}$. These calculations are presently underway. Theoretically determined *ab-initio* potential energy surfaces have already been determined for the Li + $H_2$ system [16].

Other gases contribute as minor constituents of the background or as calibration or process gases, such as $N_2$, Ar, $H_2O$, $CO_2$, $CH_4$, and He [17]. Figure 4 shows a semi-classical estimate of these rate coefficients for various sensor atoms colliding with various room temperature gases. The thermalized rates are all on the order of $10^{-9}$ $cm^3/s$ and show a 20% variation for the various constituent gases. Consequently, in the UHV/XHV environment, where typically 95% of background gas is $H_2$, uncertainty in the exact composition of the remaining gas is a small contribution to the total uncertainty. The semi-classical rate coefficients are rough estimates with unknown uncertainties, and are only used to guide our initial design. However, preliminary calculations on the Li + Li system indicate that the accuracy of the semi-classical estimates may already be better than 5 %.

Many atoms can be laser cooled and trapped, but Li or Rb are the best candidates for sensor atoms in the CAVS. The hydrogen-like atomic structure of these alkali-metal atoms makes them the easiest to cool and low-cost lasers that operate at the cooling transition wavelengths are available. Li has several advantages that make it the best choice for a primary CAVS with the lowest uncertainty. First, and perhaps most importantly, the Li + $H_2$ system has only 5 electrons and is the most theoretically tractable of the alkali plus hydrogen systems. Li also has an extremely low saturation vapor pressure, < $10^{-16}$ Pa at room temperature, thus effectively eliminating Li as a potential background gas. Moreover, the loss-rates due to two-body and three-body scattering among Li atoms in the ground state is not significant compared to collisions with the background atoms well into the XHV region. Therefore, the primary CAVS with the best uncertainty is based on the Li + $H_2$ system. On the other hand, Rb has some advantages that may make it useful for a portable miniature CAVS and other sensors based on ultra-cold atoms. The lasers required to cool Rb are inexpensive compared to those for other alkali-metal atoms, and the hyperfine structure of Rb makes it relatively easy to cool and trap compared to Li. (Unlike the $2P_{3/2}$ state in Li, the hyperfine structure of the $5P_{3/2}$ is resolved.) The vapor pressure of Rb is quite high, ~ $10^{-5}$ Pa at room temperature, which makes loading the trap with Rb less challenging, but makes reducing the Rb background more challenging—a trade-off that makes sense for a deployable device, but not for the permanently installed instrument at NIST.

Given the choices available for sensor atoms and the number of possible background gas species, SI traceable cross sections are desired for more than just the Li + H₂ system to maximize the usefulness of the CAVS. Part of the technical challenge in developing the CAVS will be to transfer SI traceability from the known *ab initio* cross sections to those that are experimentally determined. We will experimentally determine cross sections for collisions of cold alkali-metal atoms with H₂ and many other molecules by building a dynamic expansion standard in tandem with the CAVS. This allows us to generate a known gas pressure rise in the high to extreme-high vacuum range in the CAVS. By setting a known pressure and measuring the trap lifetime, the cross sections leading to trap loss are determined. Cross sections are a fundamental property and not pressure dependent, therefore their determination can be made at pressures around $10^{-8}$ torr, where the uncertainty of the generated pressure rise will be less than 2%, and checked at lower pressures. The dynamic expansion system produces a known pressure rise $p_{DE}$ in the CAVS sensing chamber from a known flow $q$ through an orifice of known dimension [18, 19]

$$p_{DE} = \frac{q}{C_{GAS}}. \tag{2}$$

For simplification, we ignore the finite pressure difference across the orifice and the background pressure in the chamber. The latter will be small compared to the pressure generated for the cross-section measurements. In the molecular flow regime, the conductance $C_{GAS}$ for gases other than H₂ is

$C_{GAS} = \sqrt{\frac{M_{H_2}}{M_{GAS}}} C_{H_2}$ where $C_{H_2}$ is the conductance of H₂ and $M_X$ is the mass of species $X$. The known gas flow will be generated using a constant-pressure flowmeter, similar in principle to those constructed and presently operated by NIST but with lower outgassing rates [19]. The constant-pressure flowmeter (CPF) produces a known flow of gas $q$ by allowing gas to flow out a variable volume $V$ through a small constriction. The pressure $p_{CPF}$ in this volume is held constant by continually reducing the volume. Thus

$$q = p_{CPF}\dot{V}; \quad \dot{V} \equiv \frac{dV}{dt}. \tag{3}$$

Putting eqs. (2) and (3) together gives

$$p_{DE} = \frac{p_{CPF}\dot{V}}{C_{GAS}}. \tag{4}$$

To experimentally determine a loss-rate coefficient, we use the dynamic expansion system to set a vacuum pressure in the CAVS, *i.e.* $p_N = p_{DE}$. Using this technique, the loss-rate coefficient for an arbitrary gas is given by

$$k_{loss}(GAS) = \frac{\Gamma_{GAS} k_B T}{p_{DE}} = \frac{\Gamma_{GAS} C_{GAS} k_B T}{p_{CPF,GAS} \dot{V}_{GAS}}. \tag{5}$$

We now turn our discussion to the concept of the Quantum SI, and consider in depth the question of whether or not the CAVS qualifies as a primary standard in this emerging paradigm. If the CAVS were solely based on an *experimentally* determined loss-rate coefficient measured as in eq.(5), the primary nature of the CAVS would be impeachable, because the measurement would be dependent on a calibrated gauge (a capacitance-diaphragm gauge). The uncertainty of the CAVS would then be correlated to that of the pressure calibration factor determined by the NIST ultrasonic interferometric manometer [20]. On the other hand, we rely on a loss-rate coefficient (for the Li + $H_2$ system) determined from *ab initio* calculations, $k_{loss}(H_2;ab)$, and we extend the primary SI traceability from the Li + $H_2$ system to Li + GAS or Rb + GAS using experimentally determined relative sensitivity coefficients,

$$S_{GAS} \equiv \frac{k_{loss}(GAS)}{k_{loss}(H_2)} = \frac{\Gamma_{GAS}}{\Gamma_{H_2}} \frac{C_{GAS}}{C_{H_2}} \left(\frac{\dot{V}_{H_2}}{\dot{V}_{GAS}}\right) \left(\frac{p_{CPF, H_2}}{p_{CPF, GAS}}\right) . \quad (6)$$

The right hand of Eq. (6) determines $S_{GAS}$ using the dynamic expansion system and CPF. Equation (6) does not depend on the absolute calibration of the pressure sensor provided the same sensor, on the same scale, is used for the measurements of both $\Gamma_{GAS}$ and $\Gamma_{H_2}$. The non-linearity of the pressure sensor scale will contribute to the uncertainty, but the absolute calibration of the sensor is not required. Therefore, *the determination of the CAVS sensitivity coefficient does not depend on a calibrated pressure sensor*. It should be noted that in our flowmeter design, the pressure sensors will be maintained at the same temperature as the flowmeter; therefore, there will be no mass-dependent thermal transpiration effect. For most gases, it is possible and desirable to set the flowmeter to the same pressure as used to determine $\Gamma_{H_2}$, i.e. $p_{CPF, GAS} = p_{CPF, H_2}$. Moreover, in the molecular flow regime the conductance will scale as the square root of the mass, and we can set $\dot{V}_{H_2}$ and $\dot{V}_{GAS}$ such that $\sqrt{M_{H_2}}\dot{V}_{H_2} = \sqrt{M_{GAS}}\dot{V}_{GAS}$ and eq. (6) simply becomes $S_{GAS} = \Gamma_{GAS}/\Gamma_{H_2}$. In any case, when the assumption of molecular flow is no longer exactly valid, $C_{GAS}/C_{H_2} \approx \sqrt{M_{H_2}/M_{GAS}}$ and deviations represent a small correction to $\Gamma_{GAS}/\Gamma_{H_2}$ for the gases of interest over the pressure range of the CPF [21,22].

Finally, with $p_{CAVS} = p_{DE}$ and combining equations 1 and 6 with the ideal gas law $p_{CAVS} = \rho_{GAS} k_B T$ the measurement equation for a primary CAVS based on Li sensor atoms for an arbitrary gas in the vacuum is

$$p_{CAVS} = \frac{\Gamma}{S_{GAS} k_{loss}(H_2;ab)} k_B T . \quad (7)$$

Relative sensitivity factors for Rb are defined the same way, with the loss rates measured for the Rb + $H_2$ and Rb + GAS systems. The measurement equation for a CAVS based on Rb becomes

$$p_{CAVS} = \frac{\Gamma}{S_{GAS} S_{Rb/Li} k_{loss}(H_2;ab)} k_B T . \quad (8)$$

Here, $S_{Rb/Li}$ is the relative sensitivity of the loss-rate coefficient for the Rb + $H_2$ system to that of Li + $H_2$.

## 4. Estimated uncertainty

A successful CAVS depends upon precise knowledge of how the loss rate of atoms from the trap depends on the density of the background gas molecules. This is seen in the measurement equation for the Li system, Eq. (7), in which the main components of uncertainty in the pressure arise from $\Gamma$ and $k_{loss}$. Uncertainty in the sensitivity coefficients $S_{GAS}$ follows from the uncertainty in $\Gamma$ and so is not separately discussed. The uncertainty in $T$ will be dominated by temperature gradients—which we anticipate to be less than 300 mK across the apparatus or 0.1%—and therefore does not contribute significantly. Uncertainty in $\Gamma$ is dominated by type-A (statistical) uncertainty; the only type-B (other than statistical) uncertainty in $\Gamma$ comes from the uncertainty in our reference clock, which is negligible.

The uncertainty in $k_{loss}$ is dominated by two effects: (1) the uncertainty of the calculated collision cross section and (2) the degree to which there is a one-to-one correspondence between collisions and ejections of sensor atoms from the trap. These contributing components are discussed below. We anticipate that combined type-B uncertainties will correspond to a pressure uncertainty of approximately 5%, dominated primarily by uncertainty in the *ab initio* cross section calculation.

It is possible that a collision between a background gas molecule and cold atom deposits only a small amount of energy to the atom, insufficient to eject it from the trap. These glancing, or "quantum diffractive" collisions constitute a small percentage of the collisions for shallow traps. Our calculations (assuming a semi-classical Li + $H_2$ differential cross section and a trap depth of 100 µK) show that fewer than one in a thousand collisions will be of this type. This fraction only depends on the trap depth and the temperature of the background gas, both of which are easily measured to better than 1%. Therefore, this effect contributes an uncertainty of less than 0.001% to the pressure measurement.

Besides background gas collisions, additional atom-loss mechanisms exist. These loss mechanisms are generally caused by interactions between cold atoms and, as such, depend on the atomic density. In this class are two-body dipolar interactions [24], three-body recombination [25], and evaporative losses [26]. These mechanisms appear as non-exponential decay. For example, consider two-body dipolar loss, which modifies Eq. (1) to become

$$\frac{d\rho_s}{dt} = -\Gamma \rho_s - b \rho_s^2 \tag{9}$$

with solution

$$\frac{\rho_s}{\rho_{s,0}} = -\frac{\Gamma}{-b\rho_{s,0} + (b\rho_{s,0} + \Gamma)e^{\Gamma t}} . \tag{10}$$

We can distinguish the two loss mechanisms by fitting the decay.

These loss mechanisms not only produce non-exponential decays, but in our experiment, they will have different timescales. Here we consider a cloud $^7$Li sensor atoms trapped magnetically in the $|F = 1, m = -1\rangle$ state with a density of $10^{10}$ cm$^{-3}$. The lifetime due to background gas collisions will range from about one second at pressures of $10^{-6}$ Pa to about $10^4$ seconds at $10^{-10}$ Pa. By comparison, the average timescales of two-body dipolar loss and three-body recombination are of the order of $10^7$ s [24, 27]. Evaporative loss timescales can be made arbitrarily long by lowering the temperature of the sensor atoms with respect to the trap depth. For a ratio of temperature of 10 µK and a trap depth of $U/k_B = 100$ µK, the evaporation timescale is on the order of $10^6$ s [26]. Thus, we do not expect these other loss mechanisms to contribute substantially to our uncertainty.

Two other decay mechanisms could perhaps produce exponential loss[1]: Majorana spin flip losses [23] and noise in the trap that cause the cold atomic gas to heat. Majorana spin-flips will be suppressed in our experiment by using a Ioffe-Pritchard magnetic trap [28]. Heating processes cause loss on timescales determined by the ratio of energy input per unit time per atom, and the trap depth. For example, in a magnetic trap generated by electromagnets, a white current noise of 1% RMS over a bandwidth of 100 kHz will cause loss on a timescale of order of $10^4$ s, which corresponds to an apparent pressure of approximately $10^{-10}$ Pa. Thus, these technical limitations will ultimately determine the practical lower limit of the device. The authors are currently preparing another manuscript to model these effects; we anticipate that our models will limit any type-B uncertainties associated with these instabilities to less than 1%.

## 5. Summary and Concluding Remarks

NIST has launched a new program to create a primary UHV/XHV standard based on ultra-cold atoms: the cold-atom vacuum standard. Present standards in the UHV [37-39] are not based on fundamental quantum properties and do not also function as sensors, and there are no existing primary standards in the XHV. The CAVS will be both a primary standard and absolute vacuum sensor; it can be made small and portable. The immediate goals of the program are to build a prototype CAVS at NIST, theoretically determine the loss rate coefficients for the Li + $H_2$ system, and measure the relative sensitivity coefficients for other gases. Further goals include building a miniature CAVS and measuring relative sensitivity coefficients for a Rb-based CAVS. We have demonstrated that primary SI traceability can be transferred from the Li + $H_2$ system to the Rb + $H_2$ system, and that primary measurements on gases other than $H_2$ can be made using sensitivity coefficients. The CAVS program fits well within NIST's Quantum SI vision, the re-definition of the SI, and the long-term goal to create other metrological quality sensors based on cold-atom technology. The CAVS will be the second primary standard to be based on cold-atom technology; the first is the fountain clock [40].

## Acknowledgements

The authors would like to thank Kirk Madison at the University of British Colombia and James Booth at the British Colombia Institute of Technology for discussions about their own cold atom sensor, helping to launch the NIST effort.## References

[1] Grimm R, Weidemüller M and Ovchinnikov Y B 2000 Optical dipole traps for neutral atoms *Adv. Atom. Mol. Opt. Phy.* **42**, 96

[2] Bergeman T, Erez G and Metcalf H J 1987 Magnetostatic trapping fields for neutral atoms *Phys. Rev. A* **35**, 1535---

[1] While these loss mechanisms are known to exist, the functional form of the associated decay is theoretically unknown.

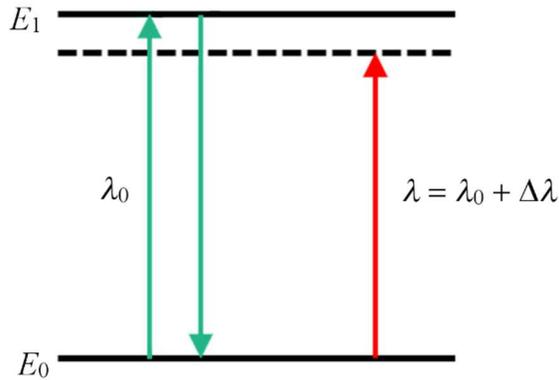

**Figure 1. Schematic diagram of energy levels in a two-state atom, resonant photons $\lambda_0$, and a red-detuned photon $\lambda$.**

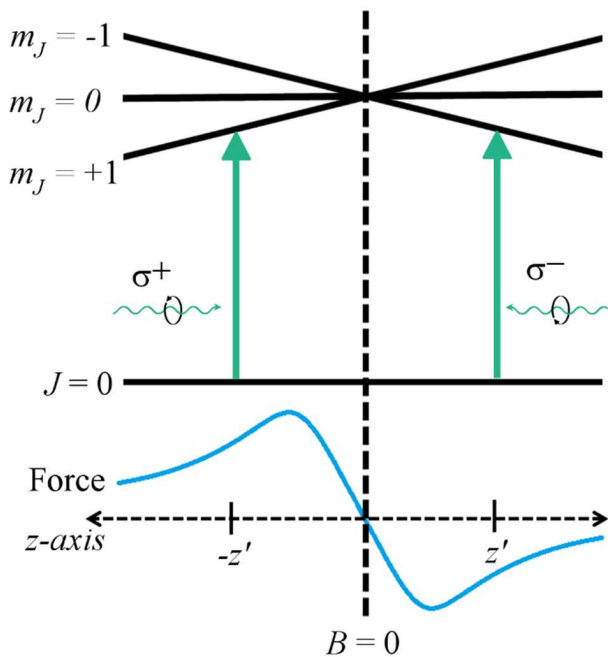

**Figure 2. Principles of the magneto optical trap. (Top) Atomic energy levels as a function of position $z$ in an applied magnetic field gradient with $B = 0$ at $z = 0$. (Bottom) This results in a position-dependent force.**

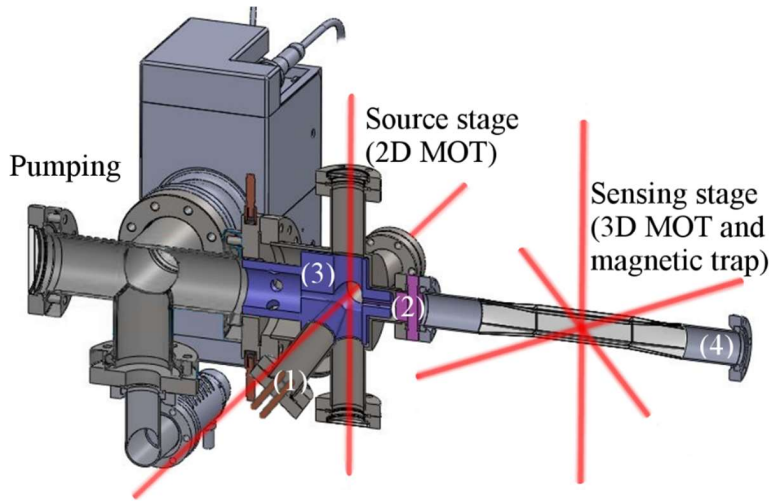

**Figure 3. Cutaway view of CAVS chamber showing the pumping region, the source stage, and the sensing stage. The alkali-metal atoms are injected by the (1) source which is angled to minimize contamination in the sensing stage while still giving the atoms some initial momentum in the axial direction. A differential pumping tube (2) allows for lower pressures in the sensing chamber. The (3) cooling shroud increases the surface sticking coefficient and thus decreases stray alkali contamination. The vacuum under test attaches at (4). Magnetic field coils are not shown.**

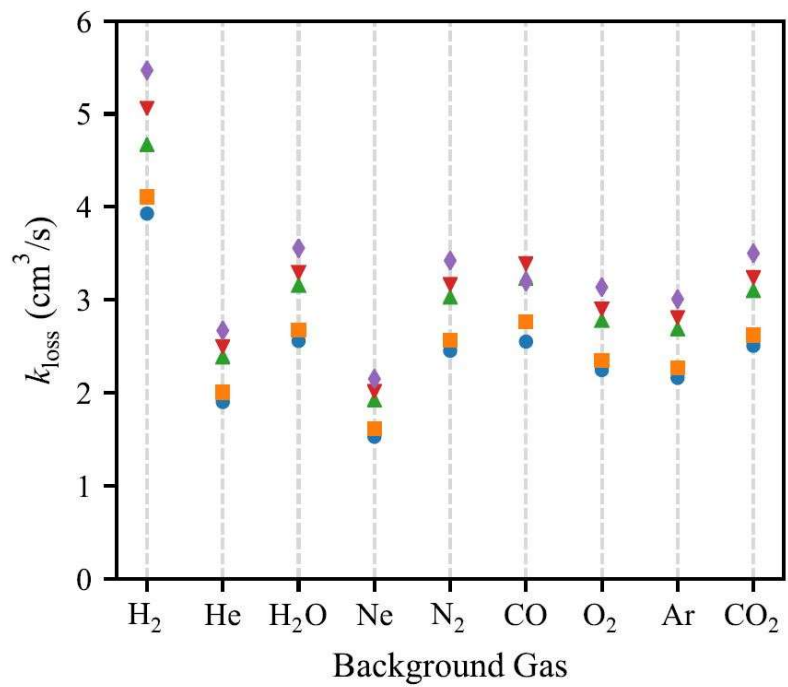

**Figure 4. Theoretical loss rate coefficient estimates, $k_{loss}$ (assuming room temperature and zero trap depth) versus background gas species for the following sensor atoms: $^7$Li (blue circles), $^{23}$Na (orange squares), $^{39}$K (inverted green triangles), $^{85}$Rb (red triangles), $^{133}$Cs (purple diamonds). These are computed using published $C_6$ coefficients where possible [29-32], or estimates based on dynamic polarizabilities and the Casimir-Polder integral [33-36].**